# Nonadiabatic exchange-correlation kernel for strongly correlated materials

Volodymyr Turkowski and Talat S. Rahman*

Department of Physics, University of Central Florida, Orlando, FL 32816, USA
*Corresponding Author, e-mail address: Talat.Rahman@ucf.edu

ABSTRACT. We formulate a rigorous method for calculating a nonadiabatic (frequency-dependent) exchange-correlation (XC) kernel required for correct description of both equilibrium and nonequilibrium properties of strongly correlated systems within Time-Dependent Density Functional Theory (TDDFT). To do so we use the expression for charge susceptibility provided by Dynamical Mean Field Theory (DMFT) for the effective multi-orbital Hubbard Model. We tested our formalism by applying it to the one-band Hubbard model: our nonadiabatic kernel leads to a significant modification of the excitation spectrum, shifting the peak that appears in adiabatic (simplified) solutions and disclosing a new one, in agreement with the DMFT solution. We also used our method to track the nonequilibrium charge-density response of a multi-orbital perovskite Mott insulator, $YTiO_3$, to a perturbation by a femtosecond (fs) laser pulse. The results were quite different from those provided by the corresponding adiabatic formalism. These initial investigations indicate that electron-electron correlations and nonadiabatic features can significantly affect the spectrum and nonequilibrium properties of strongly correlated systems.

*Introduction.*-- Correct description of the physical, including nonequilibrium, properties of strongly correlated electron systems is one of the most important goals of the condensed matter and material science communities. These systems demonstrate unusual properties with many potential applications both in bulk (high-temperature superconductivity, exotic interplay of magnetism and superconductivity, giant magneto-resistance, etc.) (see, for example Ref. [1]) and in the nanocase (for example, antiferromagnetism in small Fe chains,[2] exotic charge carrier generation in the insulating phase of a $VO_2$ nanosystem,[3] and anomalous lattice expansion[4,5] and room temperature ferromagnetism[6] in $CeO_2$ nanostructures). From the perspective of technological applications, nanostructures look even more promising than bulk, since they afford additional channels for tuning a system's properties by varying its size and geometry and by putting it on different substrates. Description of experimentally observed properties as well as prediction of new strongly correlated systems with desired properties requires reliable theoretical and computational tools.

The most successful many-body approaches for studying the static properties of strongly correlated systems — the Bethe ansatz (BA) approach[7] and DMFT[8,9] — were recently generalized to cover the nonequilibrium case (Refs. [10] and [11,12], correspondingly). The success of these theories is based on their accuracy - the Bethe ansatz solution is exact in the one-dimensional case and DMFT is very accurate in the limit of high atomic coordination number (and exact in the limit of infinite dimensions/coordination number). DMFT combined with *ab initio* DFT approaches (DFT+DMFT),[13,14] is currently being used to describe most of the types of strongly correlated systems in equilibrium: nanostructures and molecules (see, e.g., Ref. [15] and references therein), 2D systems, bulk materials,[16,17] etc. (except, probably, chains and a few-atom clusters). In the DFT+DMFT approach the "non-correlated" properties of the system, like the system geometry and spectrum, are obtained via (LDA or GGA) DFT, and the correlation effects are taken into account by post-processing the solution of the corresponding effective Hubbard model (see below). Unfortunately, DFT+DMFT is rather computationally expensive even for systems in equilibrium with dozens of non-equivalent atoms, for which computational costs are currently prohibitive beyond (geometrically and chemically) simple periodic systems or small (1nm) nanoparticles. Needless to say, application of the nonequilibrium DMFT is even more computationally expensive, especially in the case of nonhomogeneous response (e.g., domain growth) for which one has to treat all the sites as nonequivalent. For this reason, development of a pure *ab initio* TDDFT[18] for strongly correlated systems with rigorously derived XC kernel is very desirable. Indeed, being a theory of a single variable, the charge density, TDDFT allows one to study the properties of almost all types of perturbed systems of interest, from a few to thousands of atoms and to extended (periodic) systems.[19]

Some progress in this direction has already been made. In particular, Refs. [20-24] have proposed adiabatic (static DFT) XC potentials based on the BA solution. For studying time-dependent phenomena, adiabatic TDDFT XC potentials based on the results for the BA[22] and DMFT[25] XC energies for the one-band Hubbard model have also been proposed. In particular, in the last-cited paper the corresponding potential was tested versus the exact (DMFT) solution for the system in infinite dimensions. It was shown that the XC potential can reproduce the metal-insulator transition and the temporal response of systems in the case far from half-filling and moderate local Coulomb repulsions U. In other (strongly correlated) regimes the adiabatic TDDFT solution significantly deviates from the exact one, suggesting that the nonadiabatic (memory) effects might be crucial in such cases. Indeed, as we have recently shown by analyzing the exact solution for the Hubbard dimer,[26] nonadiabaticity, i.e. frequency-dependence of the XC kernel, has to be taken into account in order to obtain a correct electronic spectrum, with characteristic (Hubbard satellite) peaks due to the dynamical (time-resolved) local interactions between the electrons. Naturally, these states affect the nonequilibrum response of the system as well. A nontrivial frequency dependence of the XC kernel in the case of small clusters was also found numerically in Ref. [27]. For extended systems, 2D and 3D one-band Hubbard models, we have demonstrated that the frequency dependence of the XC kernel is also required in order to reproduce the correct spectrum with its characteristic zero-energy quasi-particle peak.[26]

In this Letter, we formulate a rigorous approach for calculating the nonadiabatic XC kernel of strongly correlated systems based on the DMFT solution for the multi-band Hubbard model. We test it on a one-band Hubbard model (for which there is a known DMFT solution) and apply it to a three-band Mott insulator, $YTiO_3$, (with known experimental and DMFT spectra) and demonstrate that the nonadiabatic effects can play an important role in the spectral properties and nonequilibrium response of systems with strong electron correlations.

*The Nonadiabatic TDDFT+DMFT formalism.*—In the TDDFT approach, to study the physical properties of a system one needs to solve the Kohn-Sham equation for the electron wave function:

$$i\frac{\partial \psi_\sigma^l(r,t)}{\partial t} = \left[-\frac{\nabla^2}{2m} + V_{ion}(r) + V_H[n](r,t) + V_{XC}[n](r,t) + V_{exc}(r,t)\right]\psi_\sigma^l(r,t), \quad (1)$$

where $\sigma$ is the spin index, $l$ is the orbital index, and $V_{ion}(r)$, $V_H[n](r,t), V_{exc}(r,t)$ are the ion, Hartree and external (e.g., elecro-magnetic field) potentials, respectively. The remaining term in the square brackets — $V_{XC}[n](r,t)$ — is the XC potential, which takes into account all the effects of the electron-electron interactions. In the approximation of linear response, the XC potential can be expanded in terms of the linear fluctuations of the charge density:

$$V_{XC}[n](r,t) \approx V_{XC\sigma}[n](r,t=0) + \int dr'dt' f_{XC}(r,t,r't')\delta n(r',t'), \quad (2)$$

where

$$f_{XC}(r,t,r't') = \frac{\delta V_{XC\sigma}[n](r,t)}{\delta n(r',t')} \quad (3)$$

is the XC kernel. We would like to emphasize that finding the linear response solution for the XC potential (2) is a much easier task than finding the general expression for $V_{XC}[n](r,t)$, which requires minimization of the energy functional in both the charge density and the self-energy (neglected here) (see, e.g., Ref. [16]). For the majority of the purposes in studying the excitations and nonequilibrium response, the linear (including memory effects) approximation is sufficient (except, probably, the case of very strong perturbations). As we show below, the exact expression for the static part of the potential $V_{XC\sigma}[n](r,t=0)$ is not essential in the nonequiibrium case. Moreover, the static potential only leads to a change of the static Kohn-Sham spectrum which can be also obtained by using the DFT+DMFT[16,17] or other approaches. The XC potential and kernel in Eqs. (1) and (2) depend on the electron density: $\delta n(r,t) = \sum_{l,\sigma}[|\psi_\sigma^l(r,t)|^2 - |\psi_\sigma^l(r,t=0)|^2]$, so the Kohn-Sham Eqs.(1), (2) must be solved self-consistently with the last equation. Therefore, the problem reduces to finding the expression for the XC kernel (3).

To find the appropriate expression for $f_{XC}(r,t;r't')$, we will refer to the case of many-body theory. In this approach, most of the single-particle properties and the collective excitations and response can be studied by calculating the spin- and orbital-dependent single-particle Green's function

$$G_{\sigma\sigma'}^{ll'}(r,t;r',t') = -\langle T c_\sigma^l(r,t) c_{\sigma'}^{l'+}(r',t')\rangle \quad (4)$$

and two-particle susceptibility

$$\chi_{\sigma\sigma'}^{ll'}(r,t;r',t') = -\langle T n_\sigma^l(r,t) n_{\sigma'}^{l'}(r',t')\rangle, \quad (5)$$

where T is the time-ordering operator, and $c, c^+$ and n are the annihilation, creation and particle number operators with the corresponding orbital, spin, space and time indices. The knowledge of the functions (5) is especially important for TDDFT, since it defines the XC kernel. Indeed, one can find the XC kernel in terms of total charge density susceptibility

$$\chi(r,t;r',t') = -\langle T n(r,t) n(r',t')\rangle \quad (6)$$

as:

$$f_{XC}(r,r'\omega) = \chi^{-1}(r,r'\omega) - \chi_0^{-1}(r,r'\omega) \quad (7)$$

(in the frequency representation). In the last equation, $\hat{\chi}_0^{-1}$ is the inverse susceptibility in the "non-interacting" (DFT) case. Plugging the orbital-density expansion for the total charge density

$n(r,t) = \sum_{l,\sigma} n_\sigma^l(r,t)$ into Eq.(6), and using Eq. (5), one can find the expression for the total-charge susceptibility in terms of the spin-orbital susceptibilities:

$$\chi(r,t;r',t') = \sum_{l,m,\sigma,\sigma'} \chi_{\sigma\sigma'}^{ll'}(r,t;r',t'). \tag{8}$$

Thus, provided the functions $\chi_{\sigma\sigma'}^{ll'}(r,t;r',t')$ are known, one can find the total-charge XC kernel from Eq. (7) by using Eqs. (8).

To find the expression for the susceptibilities Eq. (5), we shall consider an approximate model for strongly correlated systems, the Hubbard model with the Hamiltonian:

$$H = -\sum_{i,j,l,m,\sigma} t_{ij,\sigma}^{lm} c_{i\sigma}^{l+} c_j^m + U \sum_{i,l} n_{i\uparrow}^l n_{i\downarrow}^l + (U-J) \sum_{i,l} n_{i\uparrow}^l n_{i\downarrow}^l + (U-J) \sum_{i,l} n_{i\uparrow}^l n_{i\downarrow}^l + (U-2J) \sum_{i,l,m,\sigma} n_{i\sigma}^l n_{i\sigma}^m, \tag{9}$$

where $t_{ij,\sigma}^{lm}$ are the corresponding inter-(intra-) site and hopping parameters, U is the orbital Coulomb repulsion at each site, U-J and U-2J are the corresponding inter-orbital opposite-spin and inter-orbital same-spin Coulomb repulsions. Parameter J is the exchange energy. Once one has found the single-particle Green's function (4) with desired accuracy, one can obtain the susceptibility with the same accuracy from

$$\chi_{ab}(r,r',\omega) = \int d\omega' G_{ab}(r,r',\omega+\omega') G_{ba}(r',r,\omega') \tag{10}$$

(here and below *a* and *b*, *c* and *d* stand for the orbital and spin indices). This statement is a particular case of a more general Ward-Takahashi theorem, which establishes the relations between different quantities, like the multi-particle vortices (XC kernel in our case) and the single-electron self-energy, that come from conservation laws for the system (see, for example, Ref. [28]).

*DMFT.*--To calculate the Green's function, we use the DMFT approximation. In DMFT, one neglects the momentum-(space-) dependence of the electron self-energy, i.e. in this case the problem is reduced to the problem of an electron(s) on an atom, embedded into a bath of the other electrons (more details on the DMFT approach can be found in Refs. [9,16]). The system of equations for the Green's function and two other relevant functions (we mention them below) has the following form:

$$G_{ab}(i\omega_n) = \sum_k \left[\frac{1}{i\omega_n - \hat{\varepsilon}_k - \hat{\Sigma}(\omega_n)}\right]_{ab}, \tag{11}$$

$$G_{ab}^{-1}(i\omega_n) = \mathcal{G}_{ab}^{-1}(i\omega_n) - \Sigma_{ab}(i\omega_n), \tag{12}$$

$$G_{ab}(\tau) = \int D[\psi] D[\psi^*] \psi_a(\tau) \psi_b^*(0)$$

$$\times \exp\left[-\int_0^{\frac{1}{T}} d\tau' \int_0^{\frac{1}{T}} d\tau'' \sum_{c,d} \psi_c^*(\tau') \mathcal{G}_{ab}^{-1}(\tau'-\tau'') \psi_d(\tau'') + \int_0^{\frac{1}{T}} d\tau' \sum_{c,d} U_{cd} n_c(\tau') n_d(\tau')\right], \tag{13}$$

where $\omega_n = \pi T(2n+1)$ (T – temperature, n – integer number) and $\tau$ are the Matsubara frequency and time, correspondingly. The elements of the local Coulomb interaction matrix $U_{cd}$ in the last equation can be easily identified from the last three terms of the Hamiltonian Eq. (9).

In Eq.(11), the hats over the free electron energy $\hat{\varepsilon}_k$ (k - momentum) and the self-energy $\hat{\Sigma}$ functions are put to emphasize that they matrix functions. In DMFT, it is much easier to solve the equations in Matsubara representation comparing to the real-frequency calculations. Once the solution for the functions in Matsubara representation has been found, one can obtain the corresponding real-frequency dependencies by performing the analytical continuation $i\omega_n \to \omega + i\delta$. Equations (11)-(13) have the following meaning. Eq. (13) is the expression for the local (in site index) Green's function in terms of the matrix elements of the free electron energy and the interacting electron self-energy matrices $\hat{\varepsilon}_k$ and $\hat{\Sigma}(\omega)$. The electron self-energy is momentum-independent in DMFT, which allows one to obtain the exact numerical solution for the Green's function. Also, in this approach it is assumed that the Green's functions $G_{ab}(\omega)$ and self-energies $\Sigma_{ab}(\omega)$ in the many-site problem are the same as the corresponding functions in the one-site (impurity) case. Therefore, in the single-site problem one can write down the Dyson Eq.(12) that connects the Green's function and the self-energy with the "one-site" dynamical mean-field Green's function $\mathcal{G}_{ab}(\omega)$, which describes the effects of the interaction of other electrons with the impurity electron. Finally, Eq. (13) is the path-integral expression for the impurity Green's function $G_{ab}(\tau)$ defined by given dynamical mean-field and the Coulomb repulsion parameters; $\psi_a$ and $\psi_a^*$ are fermionic Grassman fields (see, e.g., Ref. [9]). The last equation is the so-called impurity equation, which is actually the only non-trivial equation in the problem. To solve it, we use computationally non-expensive second-order perturbation theory (PT) approximation,[29] which results in the following expression for the self-energy matrix:

$$\Sigma_{ab}(i\omega_l) = U_{ab}n_b + U_{ab}^2 \sum_{m,n} \mathcal{G}_{aa}(i\omega_n)\mathcal{G}_{aa}(i\omega_m)\mathcal{G}_{aa}(i\omega_n + i\omega_m - i\omega_l), \tag{14}$$

where the first (linear) term corresponds to the static (Hartree) approximation, while the second, frequency-dependent, term describes non-trivial dynamical effects that play an important role in the behavior of many strongly correlated systems. As we show below, the PT solution is sufficient for the correct description of the spectral properties of the systems of interest, though the exact solution of the impurity problem, for example the one based on the Quantum Monte Carlo (QMC)[30] or Continuous-Time QMC[31] algorithms, can be also obtained.

Since DMFT is a "spatially local" theory, one can following its successes choose the TDDFT+DMFT XC kernel to be also local in space. For farther simplification, one can assume that the kernel is separable for the space and frequency indices, and neglect the spatial dependence of the kernel:

$$f_{XC}(r,r'\omega) = A\delta(r-r')\tilde{f}_{XC}(\omega), \tag{15}$$

where $\tilde{f}_{XC}(\omega)$ is the frequency-dependent (dynamical) part obtained from Eq. (7) by using the local susceptibilities (at r=r'). In the last equation, the spatial part of the kernel $A\delta(r-r')$ (A is a parameter) is a simple static local XC kernel. This approximation can be defined as the dynamical impurity approximation, since in order to obtain the kernel from Eq.(7) one needs the local susceptibilities (r=r') which on their turn are defined by the local (impurity) Green's functions in Eq. (10). This approximation corresponds to the case when only the local charge dynamics is taken into account, which may be regarded as a reasonable choice in the case of localized d- and f-orbitals. General (spatial- or momentum-dependent) expression for the DMFT XC kernel can be also obtained from Eq. (7) by using exact susceptibilities Eq.(10). In particular, in the case of translational-invariant systems, when one can get the momentum-dependent kernel from $f_{XC}(k,\omega) = \chi^{-1}(k,\omega) - \chi_0^{-1}(k,\omega)$, where the momentum-dependent susceptibility can be obtained from the Fourier transformed spatial susceptibilities Eq.(10) by using the momentum-dependent Green's functions $G_{ab}(k,\omega)$, defined by the expression under the sum on the right hand side of Eq. (11). Below, in applications of the TDDFT+DMFT formalism we use the kernel Eq. (15).

*Applications: one-band Hubbard model.*--To test the formalism, we compare the results for the excitation spectrum (response function) of the infinite-dimensional one-band Hubbard model at half-filling (average number of electrons per site n=1) obtained with the nonadiabatic TDDFT+DMFT approach (XC kernel (15), A=1) with the results for DMFT spectral functions. As it follows from Fig.1, the two solutions are in agreement. In particular, one can reproduce the main features of the DMFT excitation spectrum with TDDFT+DMFT: the excitations between the two Hubbard bands centered around $\omega = \pm U/2$ and the low-energy intra-band excitations that correspond to the transitions inside the half-filled central band (corresponding to the "zero energy quasi-particle peak", which is suppressed at large values of U[9,32,33]). For example, at U=2t (t is the hopping parameter) the TDDFT response function (Fig. 1, right; solid green line) demonstrates these two types of excitations by peaks centered around $\omega = 0$ and $\omega = U = 2t$. The central peak is characteristic feature of the DMFT solution, which is missed in the static mean-field (MF) approximations. As a consequence the MF calculations are not capable produce the low-energy excitation peak when implemented into TDDFT.

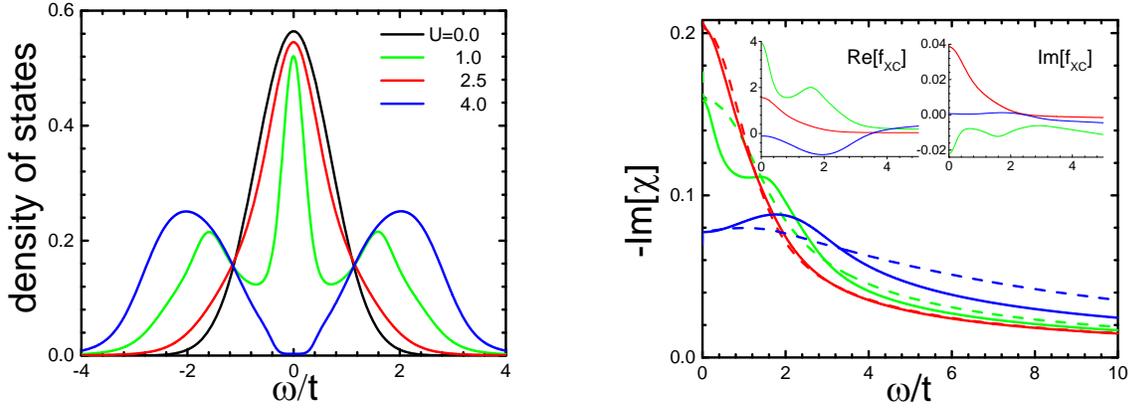

Fig. 1. Left: The DMFT DOS for the infinite-dimensional one-band Hubbard model at half-filling at different values of the Coulomb repulsion. Right: the corresponding TDDFT+DMFT imaginary parts of the response functions in the case of IPT (solid lines) and "static", or unrestricted Hartree-Fock, (dashed lines) approximations for the self-energy. The inset graph shows the frequency dependencies of the XC kernel.

To demonstrate this, we compare the full TDDFT+DMFT response function (based of the PT solution for the self-energy Eq. (13)) with the corresponding TDDFT+DMFT (or rather, TDDFT+U) results when the static (linear in U, corresponding to the unrestricted Hartree-Fock/"DFT+U") approximation for the electron self-energy was used. In the most interesting and already discussed case of an intermediate value of Coulomb repulsion, U=2t, the results for both response functions are similar at large frequencies solid and dashed green lines in Fig. 1, right), while at the intermediate $\omega's$ (between approximately t and 2t), the static approximation misses the features coming from the dynamical effects. In order to get a better understanding of this difference, one can analyze the expression for the exact analytical solution for the spin susceptibility at large U's: $\chi_{\sigma\sigma}(\omega) = \frac{Un_{\bar{\sigma}}(1-n_{\bar{\sigma}})}{\omega^2-U^2+i\delta}$, which corresponds to the static case. As it follows from the last equation, the response function in the static case demonstrates only the Hubbard band excitation with energy U, while the other peaks in the spectrum are missed. Thus, rigorous inclusion of the dynamical effects captured by DMFT is important in the TDDFT case as well.

*Applications: YTiO$_3$.*--As a multi-orbital application, we consider a Mott insulator system of YTiO$_3$ with the energy gap 1eV and the ferromagnetic critical temperature 30K.[34-36] DFT calculations[34] predict a metallic ground state with three partially occupied narrow t$_{2g}$ bands around the Fermi level with approximate total electron occupation number $n \approx 1$. Due to a weak, $\sim 0.1 eV$, splitting of the t$_{2g}$ bands (most probably caused by the GdFeO$_3$-type distortion), three orbitals (d$_{xy}$, d$_{xz}$ and d$_{yz}$) contribute to the ground state wave function. The other d-bands (E$_g$) are significantly separated in energy from the Fermi level due to the crystal field splitting. Contrary to the DFT results, DMFT calculations[34-35] and experimental studies[36] show that YTiO$_3$ is a Mott insulator. Our DMFT result for the spectral function is in agreement with these conclusions (Fig.2 left, the details of the calculations are given in the caption). It should be noted that we use the second-order PT expression for the self-energy (13) in the calculations, which can be regarded as a sufficient approximation for our purpose, while more accurate solvers are required when, for example, one is interested in detailed changes of the magnetic order in the system crossing the metal-insulator transition point (see, for instance, Ref. [34]).

Finally, we applied the TDDFT+DMFT approach to study the excited charge dynamics in YTiO$_3$ in the case of an applied external fs laser pulse (applied only to the d$_{xy}$ subsystem). We have calculated the time-dependent orbital occupancy from the Kohn-Sham equation (1) by expanding the wave function in terms of the t$_{2g}$ orbitals in the case of applied pulse $V(t) = V_0 \times \exp(-t^2/\tau^2)$ ($V_0 = 0.001$eV and $\tau = 2.05$fs are the magnitude and duration of the pulse). As it follows from our calculations (Fig. 2, right), the electron interaction effects play a very important role in the response. Indeed, in the non-interacting case only the xy-band becomes excited and with consequent relaxation to a saturated excited charge-density regime. The mean-field (Hartree-Fock) approximation predicts much stronger response with larger excited charge densities, including excitations to the other two bands due to the scattering effects. While a more general dynamical solution (Eq.(13)) is rather close to the MF one at short times, it demonstrates a much stronger response at longer times. Indeed, one may expect the memory (time, or frequency) effects to affect the post-pulse dynamics, in particular due to fluctuations in the orbital occupancies. It is important to emphasize that the second-order PT expansion is capable to correctly describe the dynamics of the system only for times below $\sim t/U^2$ [37] (femtoseconds in our case). Analysis of the dynamics at arbitrary long times requires the exact expression for the self-energy, which is beyond the scope of this paper and is planned to be performed in near future.

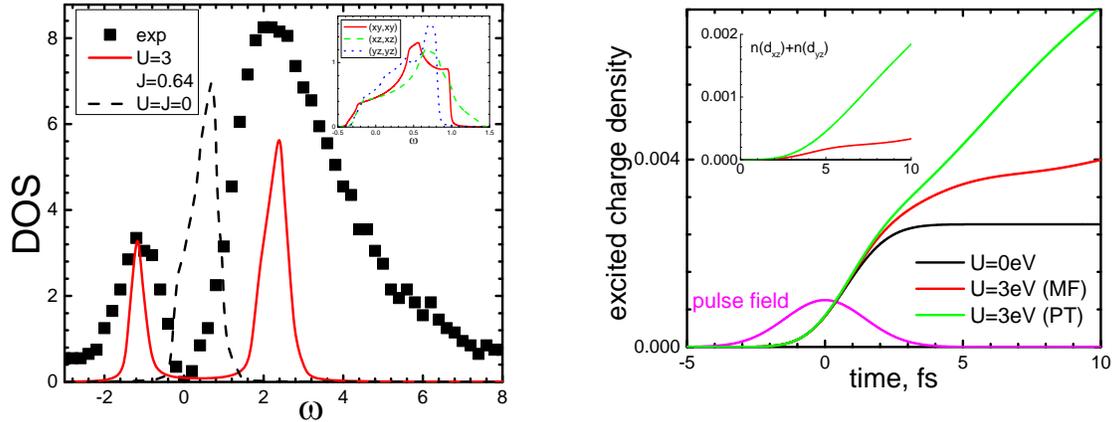

FIG. 2: Left: The DFT (U=0), DMFT and the experimental results[36] for the spectral function for the $t_{2g}$ electrons for $YTiO_3$. The insert shows the projected non-interacting spectral functions for the three bands. Right: the fs TDDFT+DMFT dynamics of the pumped charge density after a laser pulses excitation in the case of different approximations. It was assumed that the pulse is applied to the xy-band only. Insert: excited charge density for other two bands. The DMFT results were obtained from the effective tight-binding three-band Hubbard Hamiltonian with the hopping parameters obtained from DFT calculations (Ref. [34], Table V). The other parameters are U = 3eV; J = 0.64eV . A separable XC kernel (15) was used with parameter A=0.01.

*Summary.*--We have formulated a multi-orbital nonadiabatic TDDFT+DMFT approach to study the spectral properties and nonequilibrium response of strongly correlated systems. The expression for the key element of the theory - nonadiabatic XC kernel - is derived by using the DMFT solution for the electron self-energy for the effective Hubbard model. The general steps of the algorithm can be summarized in the following scheme: DFT $\to$ DMFT($\Sigma$) $\to f_{XC} \to$ TDDFT, i.e. a "non-correlated" DFT solution is used to obtain the DMFT XC kernel, which is farther implemented into the TDDFT framework to study the excitations and the nonequilibrium response of the system. We tested the formalism in the case of infinite-dimensional one-band Hubbard model at half-filling, where an accurate DMFT solution is available, and in the case of multi-orbital real material with available experimental data. It is demonstrated that the nonadiabatic effects are important for both the excitation spectrum and the response of the system. Since the DMFT approach is now regarded as almost the method of choice in the many-body strongly correlated community, we believe that the corresponding XC kernel can be successfully used in the time-dependent *ab initio* applications as well. On the other hand, farther analysis of the accuracy of the approximation is needed. The most important questions in this regard are i) inclusion of the screening/spatially-extended interaction effects (with a possible attempt to combine the DMFT and GW contributions to $f_{XC}$ (see, e.g., Ref. [38], where such an approach was employed in the equilibrium case)) and ii) the double counting correction, i.e. consistent inclusion of the correlation effects with DMFT(+GW) approximation already at the initial (relaxation, DFT) stage.[39] These and some other questions are planned to be considered in the nearest future.

*Acknowledgements.--* V.T. would like to thank E.K.U. Gross an enlightening critical discussion. We thank Lyman Baker for critical reading of the manuscript and DOE for a partial support under grant DOE-DE-FG02-07ER46354.

# REFERENCES

1. V. Anisimov, Yu. Izyumov, *Electronic Structure of Strongly Correlated Materials* (Springer, 2010).
2. S. Loth, S. Baumann, C.P. Lutz, D.M. Eigler, and A.J. Heinrich, Science **335**, 196 (2012).
3. M. Nakano, K. Shibuya, D. Okuyama, T. Hatano, S. Ono, M. Kawasaki, Y. Iwasa, and Y. Tokura, Nature **487**, 459 (2012).
4. S. Tsunekawa, K. Ishikawa, Z. Li, Y. Kawazoe, and A. Kasuya, Phys. Rev. Let. **85**, 3440 (2000).
5. R.K. Hailstone, A.G. Di Francesco, J.G. Leong, T.D. Allston, and K.J. Reed, Journ. of Phys. Chem. C **113**, 15155 (2009).
6. A. Sundaresan, R. Bhargavi, N. Rangarajan, U. Siddesh, and C.N.R. Rao, Phys. Rev. B **74**, 161306 (2006).
7. E.H. Lieb and F.Y. Wu, Phys. Rev. Lett. **20**, 1445 (1968).
8. W. Metzner and D. Vollhardt, Phys. Rev. Lett. **62**, 324 (1989).
9. A. Georges, G. Kotliar, W. Krauth, M.J. Rozenberg, Rev. Mod. Phys. **68**, 13 (1996).
10. P. Mehta, N. Andrei, Phys. Rev. Lett. 96, 216802 (2006).
11. J.K. Freericks, V.M. Turkowski, V. Zlatic, Phys. Rev. Lett. **97**, 266408 (2006).
12. H. Aoki, N. Tsuji, M. Eckstein, M. Kollar, T. Oka, P. Werner, Rev. Mod. Phys. **86**, 779 (2014).
13. V.I. Anisimov, A.I. Poteryaev, M.A. Korotin, A.O. Anokhin, G. Kotliar, J. Phys.: Condens. Matter **9**, 7359 (1997).
14. A.I. Lichtenstein, M.I. Katsnelson, Phys. Rev. B **57**, 6884 (1998).
15. V. Turkowski, A. Kabir, N. Nayyar and T. S Rahman, J. Chem. Phys. **136**, 114108 (2012).
16. G. Kotliar, S.Y. Savrasov, K. Haule V. S. Oudovenko, O. Parcollet,. C.A. Marianetti, Rev. Mod. Phys. **78**, 865 (2006).
17. K. Held, I.A. Nekrasov, G. Keller, V. Eyert, N. Blumer, A. McMahan, T. Pruschke. V.I. Anisimov, and D. Vollhardt, Phys. Stat. Sol. (b) **243**, 2599 (2006).
18. E. Runge and E.K.U. Gross, Phys. Rev. Lett. **52**, 997 (1984).
19. C.A. Ullrich, *Time-Dependent Density-Functional Theory: Concepts and Applications* (Oxford University Press, 2012).

20. N.A. Lima, M. F. Silva, L. N. Oliveira, K. Capelle, Phys. Rev. Lett. **90**, 146402 (2003); N.A. Lima, L.N. Oliveira, and K. Capelle, Europhys. Lett. **60**, 601 (2002).

21. G. Xianlong, M. Polini, M.P. Tossi, V.L. Campo, Jr., K. Capelle, M. Rigol, Phys. Rev. B **73**, 165120 (2006).

22. C. Verdozzi, Phys. Rev. Lett. **101**, 166401 (2008).

23. W. Li, G. Xianlong, C. Kollath, and M. Polini, Phys. Rev. B **78**, 195109 (2008).

24. J.P. Bergeld, Z.-F. Liu, K. Burke, and C.A. Staford, Phys. Rev. Lett. **108**, 066801 (2012).

25. D. Karlsson, A. Privitera, and C. Verdozzi, Phys. Rev. Lett. **106**, 116401 (2011).

26. V. Turkowski, T.S. Rahman, Journ. of Phys.: Cond. Matt (Fast Track) **26**, 022201 (2014).

27. F. Aryasetiawan, O. Gunnarsson, Phys. Rev. B **66**, 165119 (2002).

28. M. Fabrizio, *Lecture notes on many-body theory* (SISSA, Italy, 2013) http://www.sissa.it/cm/wp-content/uploads/2013/11/lectures.pdf.

29. A. Georges, G. Kotliar, Phys. Rev. B **45**, 6479 (1992).

30. J.E. Hirsch and R.M. Fye, Phys. Rev. Lett. **56**, 2521 (1986).

31. E. Gull, A. J. Millis, A. I. Lichtenstein, A. N. Rubtsov, M. Troyer and P.Werner, Rev. Mod. Phys. **83**, 349 (2011)

32. R. Zitko, J. Bonca, and T. Pruschke, Phys. Rev. B **80**, 245112 (2009).

33. B. Kyung, S. S. Kancharla, D. Senechal, and A.-M.S. Tremblay, M. Civelli and G. Kotliar, Phys. Rev. B **73**, 165114 (2006).

34. E. Pavarini, A. Yamasaki, J. Nuss and O.K. Andersen, New Journ. of Phys. **7**, 188 (2005).

35. E. Pavarini, S. Biermann, A. Poteryaev, A.I. Lichtenstein, A. Georges, and O.K. Andersen, Phys. Rev. Lett. **92**, 176403 (2004).

36. M. Arita, H. Sato, M. Higashi, K. Yoshikawa, K. Shimada, M. Nakatake, Y. Ueda, H. Namatame, M. Taniguchi, M. Tsubota, F. Iga, and T. Takabatake, Phys. Rev. B **75**, 205124 (2007).

37. V.M. Turkowski and J.K. Freericks, Phys. Rev. B **75**, 125110 (2007).

38. J.M. Tomczak, M. Casula, T. Miyake, F. Aryasetiawan and S. Biermann, Europhys. Lett. **100**, 67001 (2012).

39. O. Grånäs, I. Di Marco, P. Thunström, L. Nordström, O. Eriksson, T. Björkman, J.M. Wills, Comp. Mat. Science **55**, 295 (2012).